\documentclass[11pt]{amsart}

\usepackage{placeins}
\usepackage{amsthm,amsmath,amssymb,url}
\usepackage[top=35mm, bottom=30mm, left=20mm, right=20mm]{geometry}
\usepackage{caption}
\usepackage{subcaption}
\usepackage{graphicx}
\usepackage{float}
\usepackage{booktabs}
\usepackage{tikz}
\usepackage{calc}
\usepackage{changepage}
\usepackage{lipsum}
\usepackage{svg}
\usepackage{pifont}
\newcommand{\edit}[1]{\textcolor{red}{EDIT: \texttt{#1}}} % comments author to author

\usepackage[backend=biber,sorting=nty]{biblatex}
\usepackage[colorlinks,bookmarksopen,bookmarksnumbered,citecolor=red,urlcolor=red]{hyperref}

\begin{document} 
\title{A model for reference list length of scholarly articles} 
\author{Fatemeh Ghaffari}
\address{Manning College of Information and Computer Sciences, University of Massachusetts Amherst}
\author{Mark C. Wilson}
\address{Department of Mathematics and Statistics, University of Massachusetts Amherst}

\date{\today}
\begin{abstract}
We introduce and analyse a simple probabilistic model of article production and citation behavior that explicitly assumes that there is no decline in citability of a given article over time. It makes predictions about the number and age of items appearing in the reference list of an article. The latter topics have been studied before, but only in the context of data, and to our knowledge no models have been presented. We then perform large-scale analyses of reference list length for a variety of academic disciplines. The results show that our simple model cannot be rejected, and indeed fits the aggregated data on reference lists rather well. 
Over the last few decades, the relationship between total publications and mean reference list length is linear to a high level of accuracy.
Although our model is clearly an oversimplification, it will likely prove useful for further modeling of the scholarly literature. Finally, we connect our work to the large literature on ``aging" or ``obsolescence" of scholarly publications, and argue that the importance of that area of research is no longer clear, while much of the existing literature is confused and confusing.
\end{abstract}

\date{\today}
\subjclass{}
\keywords{bibliometrics, scientometrics, growth of scientific literature}
\maketitle
%===============================================================================
%%%% Main matter
\section{Introduction} \label{s:intro}

The growth of science (which we take as a synonym for ``published scholarly findings", including in fields such as humanities and social sciences) can be measured by the number of articles published, the number of authors, the number of cited articles, etc. These have been heavily studied, and each is a  ``global" indicator that requires complete enumeration. A more ``local" indicator is the length of reference lists (that is, the number of bibliography items) in an article. The distribution, or simply a statistic such as the mean or median, can be well estimated via sampling a relatively small number of articles. Another convenient and obvious property is that the reference list of a published article remains unchanged over time.

The overall distribution of the length of reference lists is obviously related to the overall distribution of citations. If we consider the directed graph where each article is a node and there is an edge from X to Y precisely when X includes Y in its reference list, then the total outdogree euals the total indegree: in other words, the number of references over all articles equals the total number of citations to all articles. However, the number of citations to a given article increases over time, which makes reference lists more convenient for many computations. For example, details about reference list length have been used for normalization in citation analysis among other topics \cite{MaBo2016}. This distinction between \emph{synchronous} (looking at reference lists) and \emph{diachronous} (looking at citations) bibliometric analysis has been made by many authors since Line \& Sandison in the early 1970's. The term \emph{diasynchronous} has also been used to describe time series of synchronous studies.

The length of reference lists has been much less studied than the number of citations received, or the number of articles published. We recommend \cite{YiBe1991} for a review of older literature, which was mostly a collection of small-sample statistical data about reference lists in various research fields or particular journals. It is easily noticeable from the older data that papers from many decades ago tended to include fewer references than those written today. More refined analysis shows that the number of references has been increasing, that there is considerable variation between research fields in the mean or median reference list length, and that there is much variation between article types (for example, review articles tend to include more references than original research articles). More recent and detailed data has been compiled \cite{ULRS+2014, LaAG2007}.  

\subsection{Our contribution}
Although several researchers have modeled the growth in the number of published articles, we are not aware of any modeling of the length of reference lists --- to our knowledge all previous work has been purely descriptive based on data. We present a simple continuous-time model for article production that links quantities, and explore its behavior. The model explicitly denies any ``obsolescence" process by which older articles become less likely to be cited than newer ones. We use a large dataset of articles published in the period 2006 -- 2016 and show that that the model fits remarkably well for our purposes, despite clearly being an oversimplification.  
In particular,  we find that over the period studied, the growth in mean reference list length is well modeled by a linear function, as is the growth of the number of articles produced, while there is a clear linear relationship between the total number of articles and mean reference list length.
We also obtain a better fit than expected (although not a good fit) to the distribution of the lengths of reference lists, which is a much more stringent test of the model. This overall good fit to data gives support to the use of the model in other contexts. We also include a generalization of the model which may be useful for more refined modeling.

We connect the above work to the literature on aging and obsolescence, and give a review of some confusions in that literature.

\section{A basic model}
\label{s:model}

Some of the key factors that seem likely to increase the length of reference lists, given by previous authors (for example \cite{YiBe1991, ULRS+2014}), are:

\begin{itemize}
\item as time progresses there are simply more possible works available that a researcher could cite;
\item modern authors use electronic keyword search, making it easier to find relevant references than in the print journal era;
\item page limits have been relaxed because of online publication, allowing more space in which to cite other work;
\item changes in citation practices (for example: peer review is more stringent now, and reviewers often suggest papers to cite; courtesy citations may be more prevalent now).
\end{itemize}

%Note that  several of these assumptions might lead to an increase in the number of references per page, and not just the total number of references. 

The main likely countervailing force is the ``obsolescence" of scientific articles, which has been studied by many authors. The rapid growth in literature is often accompanied by an improvement in understanding, meaning that older references are completely subsumed by newer ones and no longer cited (certainly, Newton's original work receives very few citations today); also, authors may prize novelty and have only finite stamina for literature review. Thus older papers may become less ``citable", by which we simply mean less attractive to a citing author than newer ones. However, against this is the fact that references are often obtained via ``snowball sampling" in which a researcher reads a given article, then follows up its references, etc. This tendency would presumably give an advantage to older papers.

It is therefore very unclear what we ought to expect from the time evolution of citability. However, the above considerations lead to two clear research hypotheses. Hypothesis A states that citability of articles is constant over time. In other words, there is no clear bias either toward or away from older articles. Hypothesis B states that citability of articles declines over time, so that there is a bias toward newer articles. Hypothesis B has been assumed (after an initial increase in citability soon after publication) by many authors in the literature on aging/obsolescence of scholarly articles.

\subsection*{The model}

Consider the following model of article production. Suppose that articles are published continuously and let $P(t)$ be the total number of articles published up to time $t$, so that $P'(t)$ is the instantaneous rate of publication at time $t$. Suppose that each new article cites each previously published article with probability $q(a)$ depending only on the age $a$. In other words, the number of citations by article $X$ published at time $t$ of each article $Y$ published at time $s < t$  is a Bernoulli random variable $c_{t-s,XY}$, and for a fixed value $t-s$ these variables  are independent and identically distributed.

The simplest special case, which explicitly uses Hypothesis A, is when $q$ is independent of $a$. In other words, 
the number of citations by article $X$ published at time $t$ of each article $Y$ published at time $< t$  is a Bernoulli random variable $c_{XY}$, these variables  are independent and identically distributed, and independent of $t$. We call this ``the uniform model" below. 

The quantity $q$ in the model should be expected to vary between research fields, even in the simplest case when it is constant. Our data below shows that the number of references varies substantially between disciplines (because of cultural differences between fields in citation behavior). For example, in our experience mathematics articles are typically written fairly tersely, and citations are mostly given to work which is directly built on by the article in question. By contrast, economics articles are notorious for their long introductions with many citations --- indeed, reviewers will often reject work that does not discuss tangentially related literature. 

\subsection*{Behavior of the model}
Our first prediction is that the expected length $L(t)$ of the reference list of each article published at time $t$ is
$$
L(t) =   \int_{-\infty}^t q(t-s) P'(s) \, ds.
$$
The intuition behind this is that $P'(s)\, ds$ is the approximate number of papers published in small time interval $[s,s+ds]$, the age for such papers is approximately $t-s$, and so each is cited with probability $q(t-s)$.

Under the uniform model we have the special case
$$
L(t) = q P(t).
$$
Differentiating, we see that in this case the growth rate of $L$ is proportional to the growth rate of $P$.

It will often be convenient to start time at some $t_0 > -\infty$ (for example, because we only have data since time $t_0$). In this case we can simply rewrite to obtain
$$
L(t) = L(t_0) + \int_{t_0}^t q(t-s) P'(s) \, ds.
$$

An important feature of the model is that because we assume uniformly random choice of article to cite, if we consider the restriction whereby we count only citations to articles published at time $t_0$ or later, the analysis is essentially unchanged. For example, defining $L^*(t)$ to mean the number of references published at time $t_0$ or later, and $P^*(t) = P(t) - P(t_0)$ the number of articles published at time $t_0$ or later, we have
$$
L^*(t) = \int_{t_0}^t q(t-s) (P^*)'(s) \, ds.
$$

\section{Experimental setup}
\label{s:experiment}

We used a dataset supplied by the company Academic Analytics \url{https://aarcresearch.com/}. The dataset is rather comprehensive and covers all publications in the period 2006-2016 by faculty members at a list of 390 PhD-granting universities in North America. More details on the process used can be found in Appendix~\ref{app:methods}. We note that such a comprehensive dataset is not really necessary for much of the analysis in the present article, because random sampling will be good enough. 

We filtered out entries with missing data and those with fewer than 5 references (in order to eliminate letters to the editor). The dataset, taken as a sample of the total amount of articles published over the given time period, seems likely (to us) to be unbiased as far as reference list length goes. In this dataset we have 
%(200970, 73091, 40159, 21791)
(147149, 53776, 30160, 16957) articles published in 2006 -- 2016 from the fields of Chemistry, Mathematics, Economics and Oncology respectively, and the number of pages and number of references of each article. 

We also needed estimates of $P(t)$. For this we used manual queries in Scopus. Of course, we do not know whether the research field classification used by Scopus is the same as that used in our provided dataset. This was one reason why we restricted to the four fields listed above --- the original dataset contained several other research fields such as History and Geography, where we expected some classification inconsistencies.

All statistical analyses were performed using Python with standard packages \texttt{numpy}, \texttt{pandas}, and \texttt{matplotlib}.

\section{Results}
\label{s:results}

\begin{figure}[htbp]
     \centering
     \begin{subfigure}[b]{0.45\textwidth}
         \centering
         \includegraphics[width=\textwidth]{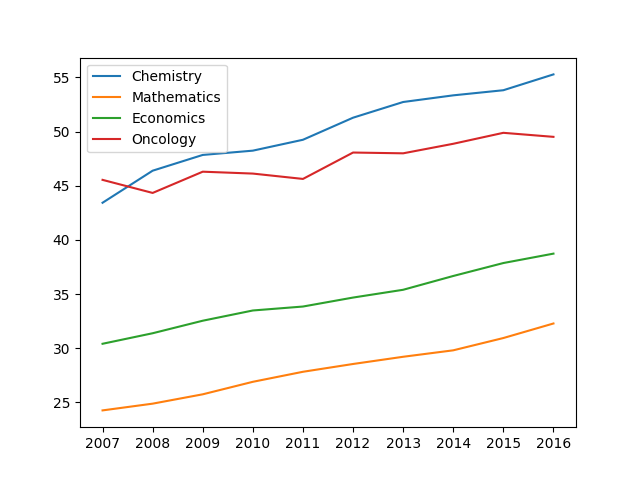}
         \caption{$L(t)$ vs $t$}
         \label{fig:LvP_L}
     \end{subfigure}
     \begin{subfigure}[b]{0.45\textwidth}
         \centering
         \includegraphics[width=\textwidth]{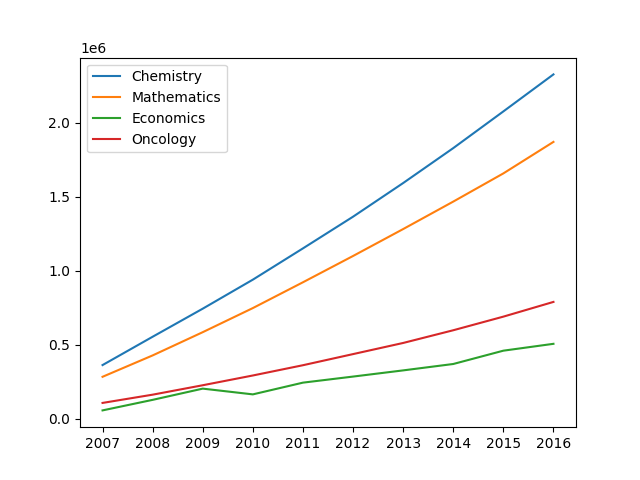}
         \caption{$P(t)$  vs $t$ }
         \label{fig:LvP_P}
     \end{subfigure}
     \begin{subfigure}[b]{0.45\textwidth}
         \centering
         \includegraphics[width=\textwidth]{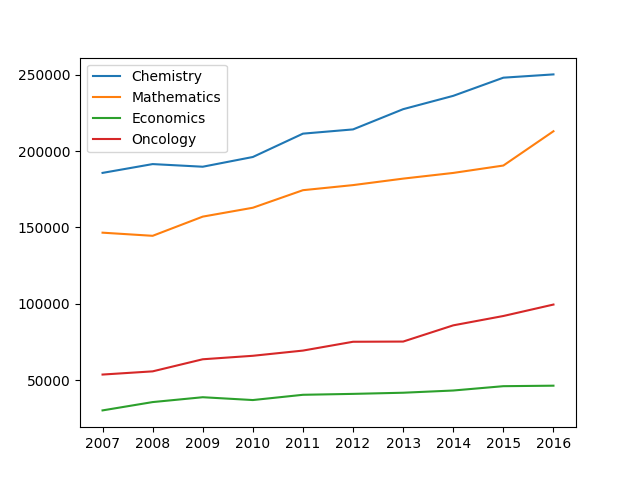}
         \caption{$P(t) - P(t - 1)$ vs $t$}
         \label{fig:LvP_PP}
     \end{subfigure}
        \caption{Relationships between $t$, $P(t)$ and $L(t)$.}
        \label{fig:LvP}
\end{figure}

In each research field under investigation, we found a steady increase  over the period 2006--2016 in the  mean number of articles published each year, and in the number of references per article.  The fit of each via ordinary linear regression was good (see Table~\ref{t:LP_fit}), suggesting a linear growth in reference list length over time. We found noticeable and consistent differences by research field in the mean reference list length, as shown in Figure~\ref{fig:LvP} (right). 

\begin{table}[htbp]
    \centering
    \begin{tabular}{c|c|c|c|c}
       Field  & $R^2$ for $L(t)$ & $R^2$ for $P(t)$ & $R^2$ for $L$ vs $P$ & $R^2$ for $P(t) - P(t-1)$\\
       \hline
       Chemistry & 0.405948 & 0.997372 &0.963810 &0.964448\\
       Mathematics & 0.605758 & 0.997737 &0.993011 &0.947762\\
       Economics & 0.521143 & 0.968208 &0.978862 &0.908734\\
       Oncology & 0.269766 & 0.993056 &0.888816 &0.969280\\
    \end{tabular}
    \caption{Fit of linear models for $L(t)$ and $P(t)$, 2006-2016 (as shown in Figures~\ref{fig:LvP} 
    and~\ref{fig:LvP_mean}).}
     \label{t:LP_fit}
\end{table}

% \begin{figure}[htbp]
% \includegraphics[width=0.45\linewidth]{pictures/Median/P_t-P_t-1.png}
% \includegraphics[width=0.45\linewidth]{pictures/Mean/L.png}
% \caption{Number of articles published (left), and mean reference list length (right), 2006-2016.}
% \label{fig:LvP}
% \end{figure}

% \begin{figure}[htbp]
% \includegraphics[width=0.4\linewidth]{pictures/Mean/Linear_C_chem_mean.png}
% \includegraphics[width=0.4\linewidth]{pictures/Mean/Linear_C_math_mean.png}
% \includegraphics[width=0.4\linewidth]{pictures/Mean/Linear_C_econ_mean.png}
% \includegraphics[width=0.4\linewidth]{pictures/Mean/Linear_C_onc_mean.png}
% \caption{Linear relationship between $P(t)$ and $L(t)$.}
% \label{fig:LvP_mean}
% \end{figure}

\begin{figure}[htbp]
     \centering
     \begin{subfigure}[b]{0.45\textwidth}
         \centering
         \includegraphics[width=\textwidth]{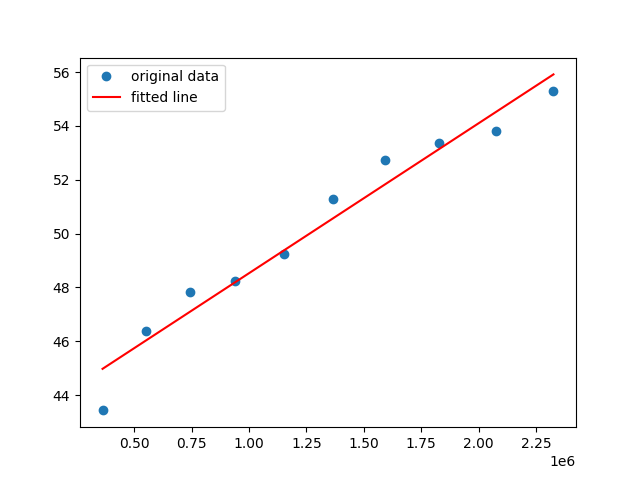}
         \caption{Chemistry -- Slope: $6\times 10^{-6}$}
         \label{fig:LvP_mean_chem}
     \end{subfigure}
     \begin{subfigure}[b]{0.45\textwidth}
         \centering
         \includegraphics[width=\textwidth]{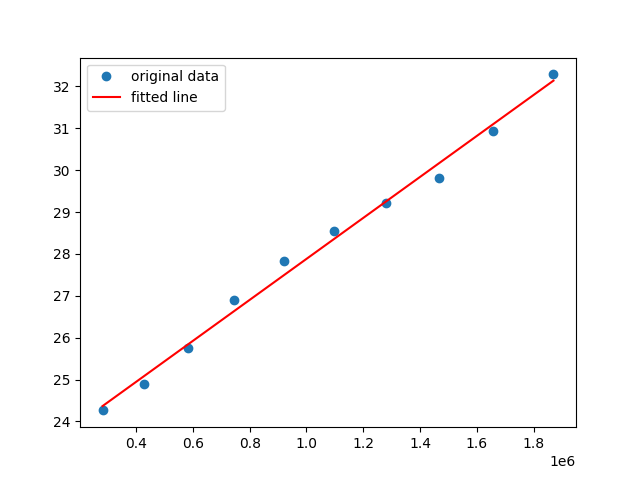}
         \caption{Mathematics -- Slope: $5\times 10^{-6}$}
         \label{fig:LvP_mean_math}
     \end{subfigure}
     \begin{subfigure}[b]{0.45\textwidth}
         \centering
         \includegraphics[width=\textwidth]{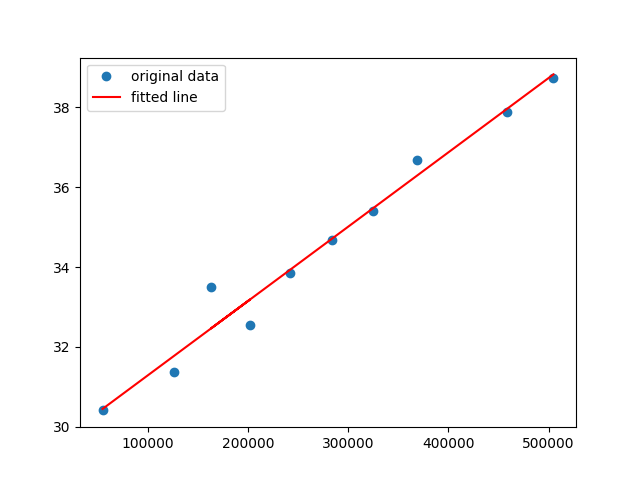}
         \caption{Economics -- Slope: $19\times 10^{-6}$}
         \label{fig:LvP_mean_econ}
     \end{subfigure}
     \begin{subfigure}[b]{0.45\textwidth}
         \centering
         \includegraphics[width=\textwidth]{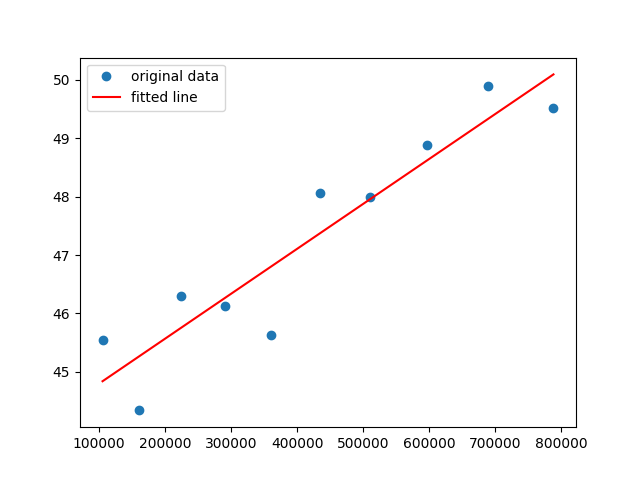}
         \caption{Oncology -- Slope: $8\times 10^{-6}$}
         \label{fig:LvP_mean_onc}
     \end{subfigure}
        \caption{Linear relationship between $P(t)$ and $L(t)$.}
        \label{fig:LvP_mean}
\end{figure}

We compared the mean list length $L(t)$ against our estimates of $P(t)$ from Scopus. The uniform case of our model predicts an affine relationship (there is a nonzero constant term coming from all articles published up to 2006) of the form $L(t) = q\left(P(t)-P(t_0)\right) + d$. The fit was extremely good (see Figure~\ref{fig:LvP_mean} and Table~\ref{t:LP_fit}). We expect that  different research cultures will lead to different values of $q$ in our model; for example we would expect the value of $q$ in the model to be larger for Economics than Mathematics, to the extent that they are disjoint fields. The slope values we obtained from the least squares fitting mentioned above indeed differed substantially, and that for Economics was much larger than that for Mathematics.

\begin{figure}[htbp]
     \centering
     \begin{subfigure}[b]{0.45\textwidth}
         \centering
         \includegraphics[width=\textwidth]{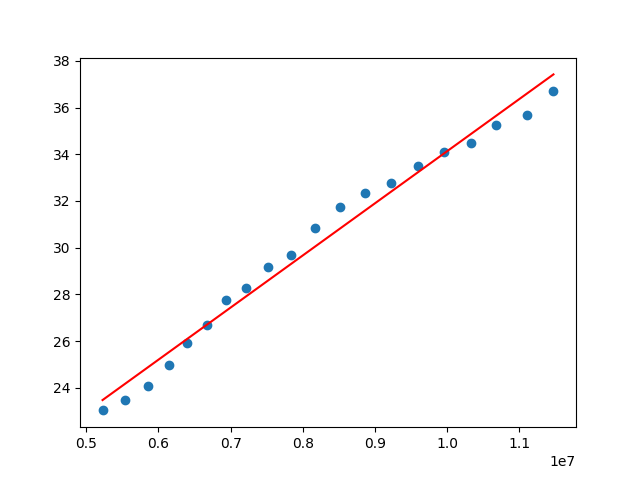}
         \caption{Medicine}
         \label{fig:LaAG2007_data_2_med}
     \end{subfigure}
    %  \hfill
     \begin{subfigure}[b]{0.45\textwidth}
         \centering
         \includegraphics[width=\textwidth]{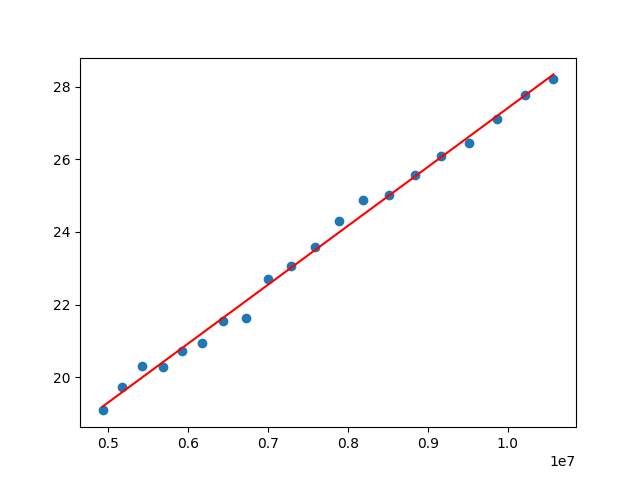}
         \caption{Natural Sciences}
         \label{fig:LaAG2007_data_2_NSE}
     \end{subfigure}
    %  \hfill
     \begin{subfigure}[b]{0.45\textwidth}
         \centering
         \includegraphics[width=\textwidth]{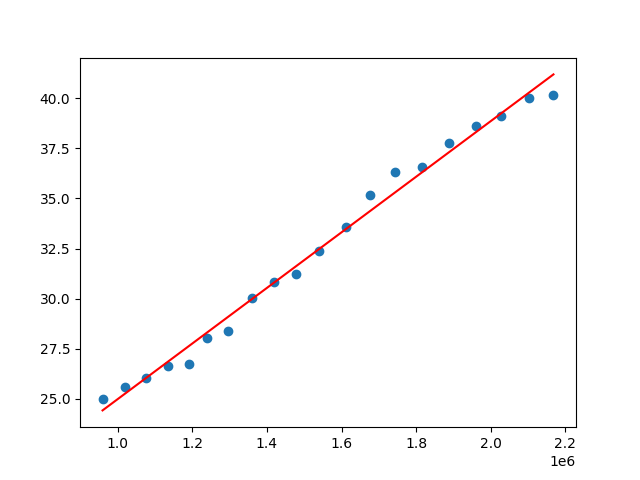}
         \caption{Social Sciences}
         \label{fig:LaAG2007_data_2_SS}
     \end{subfigure}
    %  \hfill
     \begin{subfigure}[b]{0.45\textwidth}
         \centering
         \includegraphics[width=\textwidth]{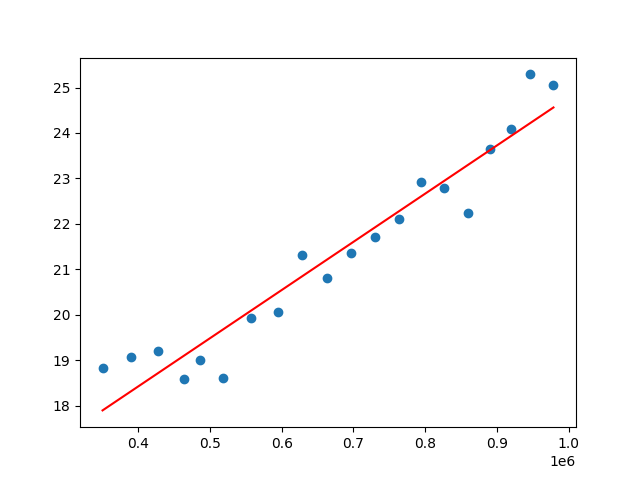}
         \caption{Arts and Humanities}
         \label{fig:LaAG2007_data_2_AH}
     \end{subfigure}
        \caption{Linear relationship between $P(t)$ and $L(t)$ from dataset of \cite{LaAG2007}, 1985--2004.}
        \label{fig:LaAG2007_data_2}
\end{figure}

We performed several robustness checks on the above results. The appendix shows how replacing the mean with the median makes relatively small changes. We also checked against another dataset. Data kindly supplied by the authors of the detailed study \cite{LaAG2007} show clearly that from around 1980 there has been a roughly linear growth in $P$ and $L$, and that $P$ is closely fitted by a linear function of $L$ (see Figure~\ref{fig:LaAG2007_data_2}).

\begin{table}[htbp]
    \centering
    \begin{tabular}{c|c}
       Field  & $R^2$ \\
       \hline
       Medicine  & 0.982209\\
       Natural Sciences & 0.994795\\
       Social Sciences & 0.990064\\
       Arts and Humanities & 0.925461\\
    \end{tabular}
    \caption{Fit of linear model for $L(t)$ versus $P(t)$ from dataset of \cite{LaAG2007}, 1985--2004.}
    \label{t:LvP_fit_old}
\end{table}

\subsection{Our model and the exponential growth phase}
\label{ss:exp}

\begin{figure}[htbp]
     \centering
     \begin{subfigure}[b]{0.45\textwidth}
         \centering
         \includegraphics[width=\textwidth]{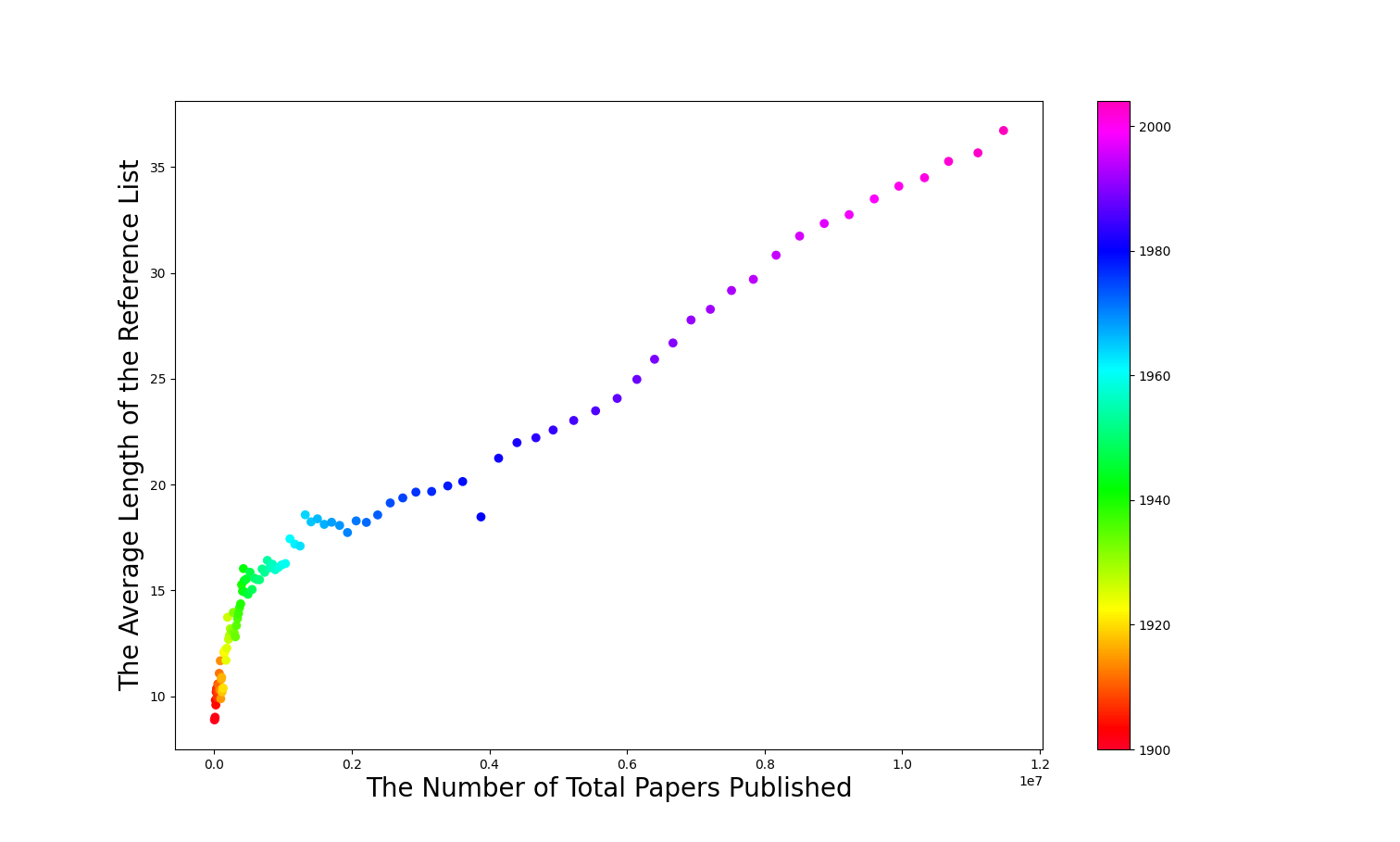}
         \caption{Medicine}
         \label{fig:LaAG2007_data_MED}
     \end{subfigure}
    %  \hfill
     \begin{subfigure}[b]{0.45\textwidth}
         \centering
         \includegraphics[width=\textwidth]{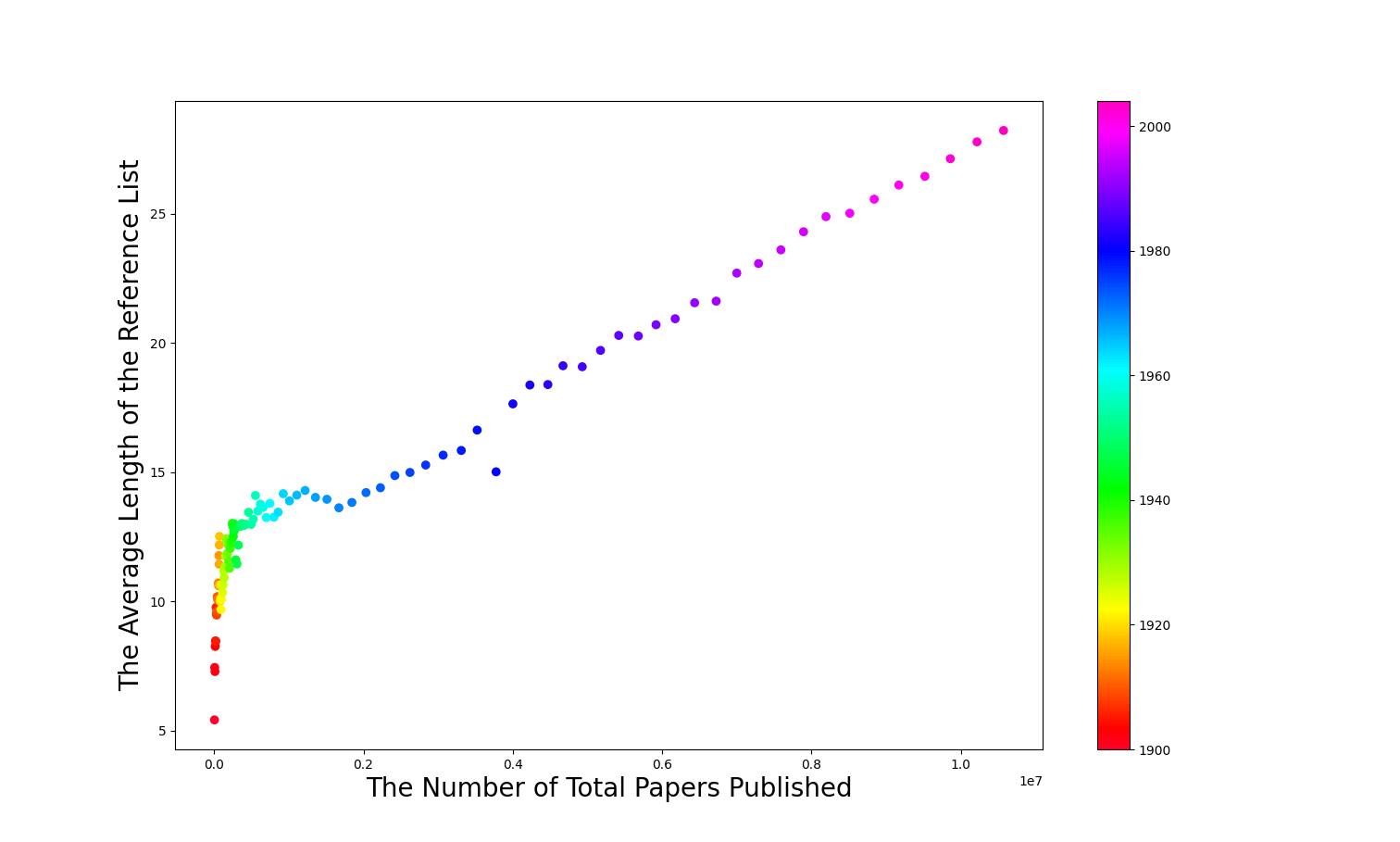}
         \caption{Natural Sciences}
         \label{fig:LaAG2007_data_NSE}
     \end{subfigure}
    %  \hfill
     \begin{subfigure}[b]{0.45\textwidth}
         \centering
         \includegraphics[width=\textwidth]{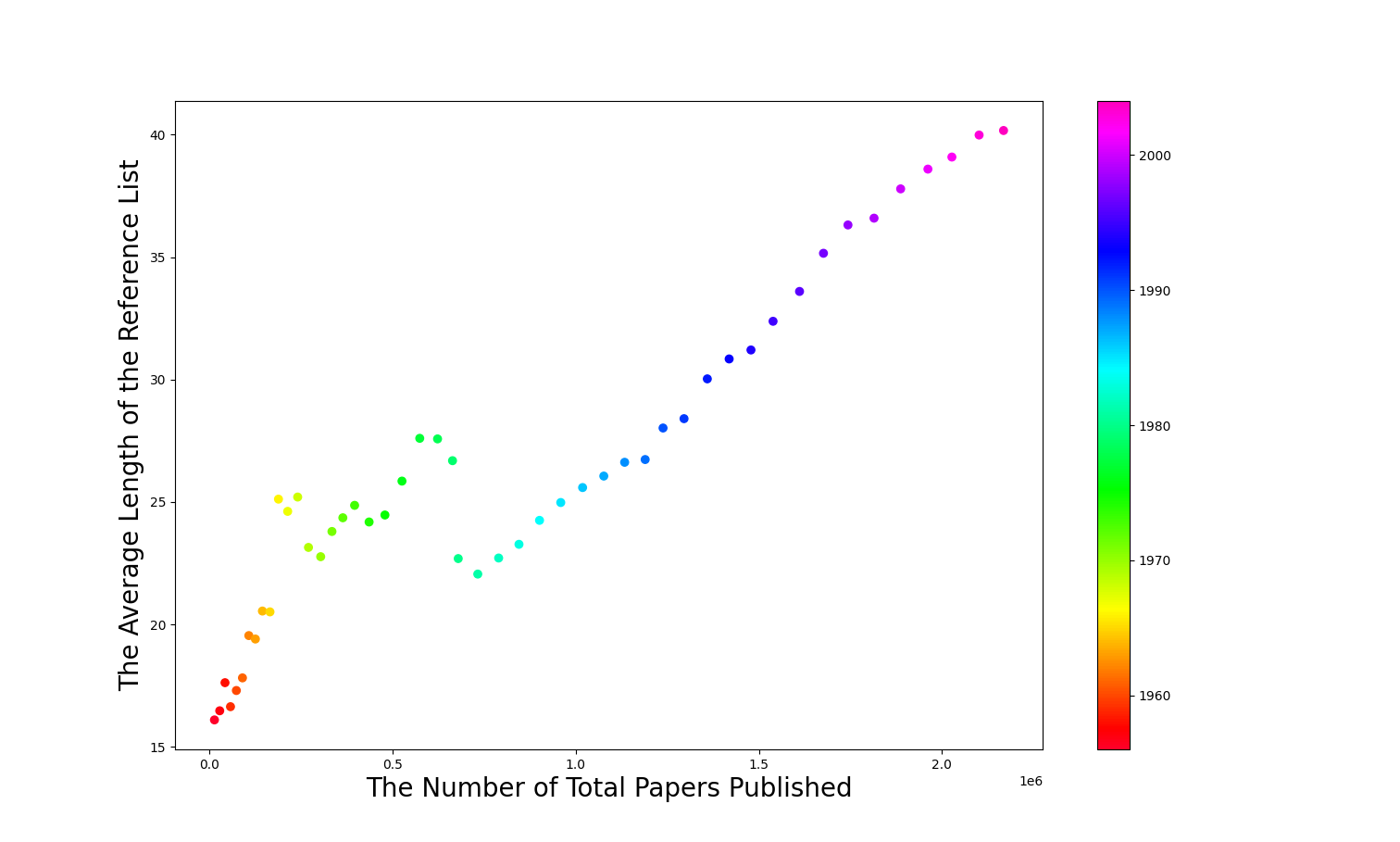}
         \caption{Social Sciences}
         \label{fig:LaAG2007_data_SS}
     \end{subfigure}
    %  \hfill
     \begin{subfigure}[b]{0.45\textwidth}
         \centering
         \includegraphics[width=\textwidth]{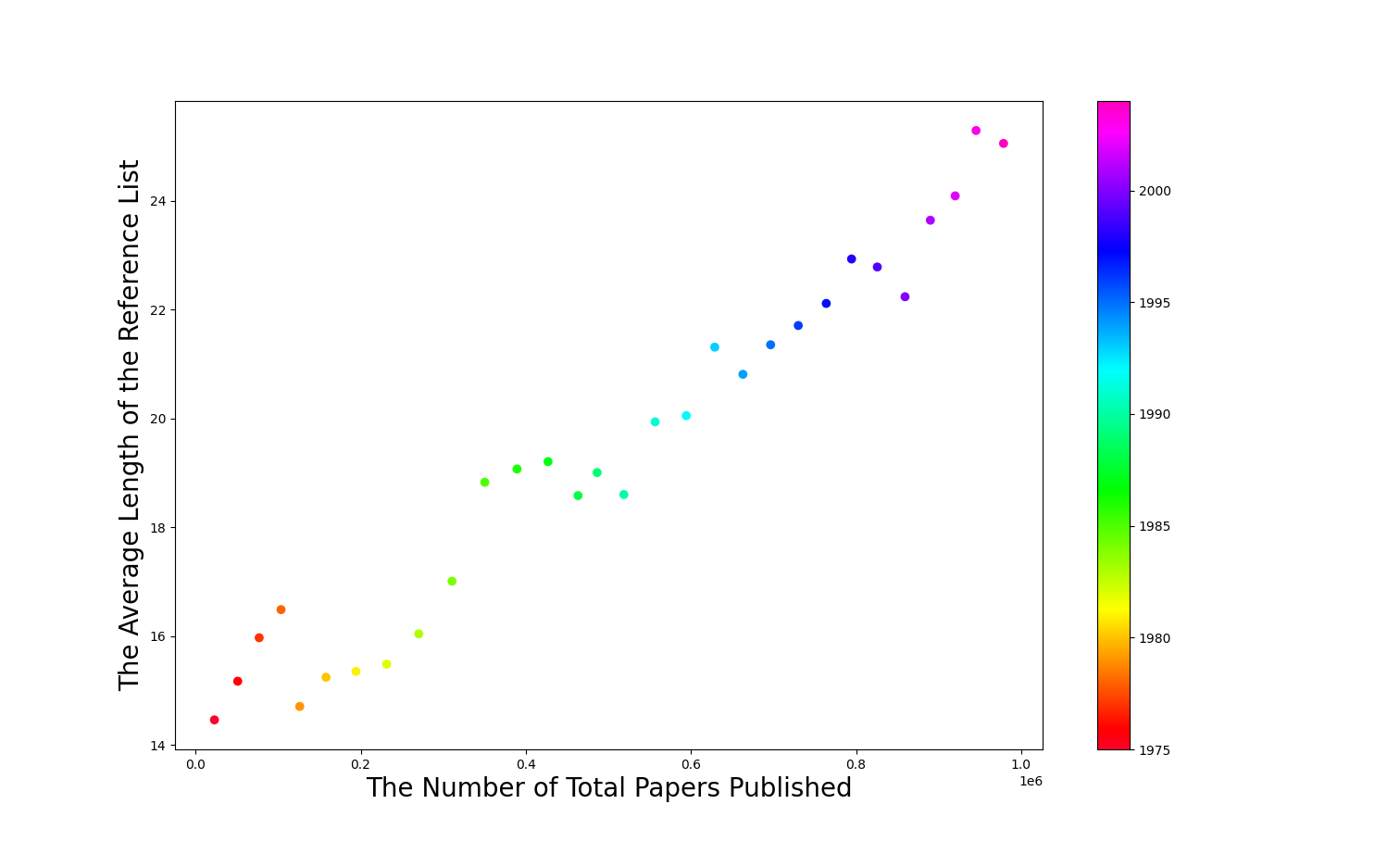}
         \caption{Arts and Humanities}
         \label{fig:LaAG2007_data_AH}
     \end{subfigure}
        \caption{Relationship between $P(t)$ and $L(t)$ from dataset of \cite{LaAG2007}, 1900--2004.}
        \label{fig:LaAG2007_data}
\end{figure}

Our uniform model predicts an exponential growth rate of reference list length during a period in which growth in articles is exponential.  The data supplied by the authors of  \cite{LaAG2007} contradicts this clearly: the growth of reference lists, the growth in $P$, and the relationship between $L$ and $P$ in various aggregated subfields over the period 1900--2004 are shown in Figure~\ref{fig:LaAG2007_data}. 

In fact no function $q$ in our more general model will yield a subexponential growth rate of reference lists when literature is growing exponentially, as the following computation shows, assuming $P(t) = Ce^{kt}$: 

\begin{align*}
L(t) & = Ck \int_{-\infty}^t q(t-s) e^{ks} \, ds \\
& = Ck \int_{0}^\infty q(a) e^{k(t-a)} \, da \\
& = Cke^{kt} \int_{0}^\infty q(a) e^{-ka} \, da. 
\end{align*}
The integral in the last line is a nonnegative constant.

These results are not surprising, since the justification for the model is lacking in this case.  It is not reasonable to assume that during the exponential growth phase of science (before the 1980s), authors could consistently locate all relevant references. For example, that time period was before the advent of widespread electronic search for publications. A more reasonable assumption is that authors added a few references to the reference lists of articles that they had read. This would lead to fairly slow growth in $L$, perhaps a little faster than linear.

\section{Connection to the literature on obsolescence and aging}
\label{ss:aging}

\subsection{Obsolescence and aging}
\label{ss:obs}

A seemingly very different topic in bibliometrics is the aging/obsolescence of scientific literature.  Having reviewed at least 30 articles from the bibliometric literature on these topics, we make several critical comments about that literature. 

First, we deal with terminology. Some authors use ``aging" as a synonym for ``obsolescence".  Obsolescence is a property of a given article (becoming less attractive to cite over time). Obsolescence occurs as the article gets older, but it is not very accurate to say that an article becomes obsolete just because it is older. Rather, it is also because more relevant, more citable articles have been published in the interim. 

Some authors have used the term ``aging" to refer to an increase over time in the (mean or median) age of items occurring in reference lists. This usage may easily lead to confusion, because increase in the age of reference lists is caused by a \emph{decrease} in the rate of ``aging" (in the sense of obsolescence) of articles! 

From now on we refer only to \emph{obsolescence of articles} and \emph{recency bias of reference lists}. The connection of our work on reference list length to these topics is as follows. 
\begin{enumerate}
\item Growth in scientific literature leads to recency bias in reference lists: older items are crowded out by the more numerous newer ones. This is true even if the length of reference lists keeps pace with the growth in article production. It is not necessary to postulate any obsolescence effect (the decline in citability of articles as time passes) in order for recency bias in reference lists to occur. 
\item However, any age-related obsolescence effect will accentuate the recency bias of reference lists. 
\item Many authors have taken obsolescence (that is, Hypothesis B) as given, and  used recency bias in reference lists as a way to estimate obsolescence.
\item Recency bias in reference lists is caused by both growth of literature and obsolescence.
\end{enumerate}

The first point holds because this behavior is seen in our uniform model, as discussed in Section~\ref{ss:age_model} below. In this model, the fraction of papers cited at time $t$ and having age at least $a$ is simply $P(t-a)/P(t)$, which decreases as $a$ increases. This obvious effect has been noted before. For example, in \cite{LaAG2007} we find: ``we suggest that the principal cause of the increased age of the cited scientific information is a fairly mechanistic response to the phenomenal growth in the quantity of published material after the war and to the current slowing in the growth of science." 

The second point is clear. If $q$ is a decreasing function of $a$, at least after a short initial period, then we would expect this expression for $L(t)$ to be smaller than when $q$ is constant.

We come now to the third point listed above. Line \& Sandison \cite{LiSa1974} point out that obsolescence is an assumption that must be verified, and there is evidence that it does happen. Larivi\`{e}re, Archambault \& Gingras \cite{LaAG2007} assume obsolescence (after, as usual with such an assumption, an initial increase in citability) as common knowledge.  After a diasynchronous study of the age of items in reference lists of articles published in the period 1900--2004, they find: ``Indeed, in contrast to a widely-held belief, scientific literature does not become obsolete faster nowadays and, actually, quite the opposite is observed. The useful life of scientific publications has been increasing steadily since the mid-seventies." This is correct insofar as an increase in the rate of obsolescence should lead to a reduction in the average age of items in reference lists, all other things being equal. However, the existence of obsolescence has not been shown by this study --- the results on recency bias in reference lists can be explained without it, as explained above.

We now address the fourth point listed above. Line \& Sandison \cite{LiSa1974} surveyed the area almost 50 years ago, and most of our observations were already noted by them or earlier authors. However, the distinction between obsolescence of articles and recency bias in reference lists, and the fact that both obsolescence and growth of the literature play a role in recency bias of reference lists, has subsequently been missed by several authors. For example Gupta \cite[p. 336]{Gupt1998} states clearly as justification for a synchronous approach to obsolescence that ``Evidence of the obsolescence of publications is presumed, if the use of these publications declines with age. Decline in the use literature over time (aging or decay) may be ascertained through studies of library use or by studying age of citations in publications or articles cited." 

Some of the confusion between obsolescence and recency bias may be caused by a mistake in conditional probability: we want to know whether $P(cited \mid old)$ is small, but we only observe $P(old \mid cited)$. Bayes' formula shows
$$
P(cited \mid old) P(old) = P(old \mid cited) P(cited).
$$
Since $P(old)$ may be considerably smaller than $P(cited)$, we cannot make the desired conclusion. 

In  \cite{LaAG2007} we find: ``we suggest that the principal cause of the increased age of the cited scientific information is a fairly mechanistic response to the phenomenal growth in the quantity of published material after the war and to the current slowing in the growth of science." Growth by itself cannot explain recency bias in reference lists, as our uniform model shows. A consequence of our model is that a slowdown in article production will lead to an increase in the average age of items in reference lists. However, if obsolescence is really a factor, then this property need not hold.

As a more general criticism, we claim that the literature on obsolescence is of rather limited practical relevance today, and some of it appears to have been misconceived from the start. As explained by Line \& Sandison \cite{LiSa1974}, we should distinguish between obsolescence of documents and of knowledge. In the age of print collections, limitations on shelf space made it important for librarians to determine which serials were less likely to be consulted in future by their patrons.  However, in the electronic age where it is possible to maintain easy access to the long tail of less popular items, obsolescence of documents is much less of a concern. It is not clear whether obsolescence of knowledge is a major problem, or ever was.

In summary, we find the literature on aging/obsolescence to be in serious need of improvement, and hope that our contribution here will spur further more rigorous work.

\subsection{The uniform model and the age distribution}
\label{ss:age_model}

Consider, under our general model, the reference list of  articles published at time $t$. The expected total age of all items in such a list is
$$
A(t):= \int_{-\infty}^t (t - s) q(t-s) P'(s) \, ds.
$$
Note that $A(t)/L(t)$ gives the mean age of an item in the reference list of a randomly selected article published at time $t$.

We restrict now to the uniform model, where $q(a)$ is constant for all $a\geq 0$. Since it will be difficult to analyse this in full generality owing to lack of historical data on $P$, we restrict to citations to articles published at time $t_0$ or later by articles appearing at time $t_0$ or later, to obtain the formula
$$
A^*(t) = q\int_{t_0}^t P^*(s) \, ds.
$$
Integration by parts allows us to rewrite this quantity as
\begin{align*}
A^*(t)& = q\int_{t_0}^t (t - s) P'(s) \, ds \\ 
& =  qt \int_{t_0}^t P'(s) \, ds -  q \int_{t_0}^t s P'(s) \, ds \\
& =  qt \left(P(t) - P(t_0)\right) - q \left[ sP(s)\right ]_{t_0}^t + q\int_{t_0}^t P(s)\, ds\\
& = q\int_{t_0}^t P(s) - q(t-t_0)P(t_0)\, ds\\
& = q\int_{t_0}^t \left(P^*(s) - P(t_0)\right) \, ds\\
& = q\int_{t_0}^t P^*(s)\, ds.
\end{align*}

\if01
Here we assume that $\lim_{s\to -\infty} sP(s) = 0$, which seems reasonable given the observed exponential growth in articles since the beginning of time until recent decades (see for example \cite{Pric1965, LaAG2007}; there is some disagreement in the literature as to whether exponential growth has ended or merely slowed, but no dispute that ).
\fi
%In particular, $$A'(t) = c \left[1 - \frac{P'(t) \int_{-\infty}^t P(s) \, ds}{P(t)^2}\right].$$

We use a ``mean-field approximation" approach, estimating the expected mean age of an item by 
\begin{equation}
\label{eq:mean age}
\frac{A^*(t)}{L^*(t)} = \frac{\int_{t_0}^t P^*(s)}{P^*(t)}.
\end{equation}

and the median by solving $P^*(t-a) = P^*(t)/2$ for $a$. Note that the expected mean age is is at most $t-t_0$, and equals the reciprocal of the fraction of the rectangle formed by the points $(t_0, P(t_0))$, $(t, P(t_0))$, $(t, P(t))$ and $(t_0, P(t))$ that lies under the graph $y = P(t)$. If for example $P^*$ grows quadratically in $t$ (corresponding to a linear increase in the number of articles produced per unit time), then these approximations yield a mean age of $(t-t_0)/3$ and median $(t-t_0)(1-2^{-1/3}) \approx 0.206(t-t_0)$. We tested \eqref{eq:mean age} on our dataset --- Table~\ref{t:age_predict} shows the predictions. To obtain data on age of items in reference lists, we randomly sampled 30 articles from each field for publication year 2016, and extracted their data from CrossRef. This required manual queries: because the details about a publication's references are not available using CrossRef's API, the entire reference list was manually extracted for each item. Afterward, a Python program was used to analyse the reference list and calculate the sample mean and median of the referenced items. Only the references published after 2006 were considered. The results are also shown in Table~\ref{t:age_predict}. The fit is good given the sample sizes.

\begin{table}[htbp]
    \centering
    \begin{tabular}{c|c|c|c|c}
       Field  & mean age (prediction) & mean age (sampling) & median age (prediction) & median age (sampling)\\
       \hline
       Chemistry & 5.36 & 4.74  & 5 & 5\\
       Mathematics &  5.40 &  5.19 & 5 & 5\\
       Economics &  4.62 &  5.37 & 5 & 5\\
       Oncology &  5.65 &  4.79 & 5 & 5\\
    \end{tabular}
    \caption{Predictions versus sample statistics for age of references conditional on article being published in 2016 and references being published no earlier than 2006.}
    \label{t:age_predict}
\end{table}

%Thus, for example, during the exponential growth phase of scientific literature, where $P(t) = Ce^{kt}$ for some positive constants $C, k$, we obtain $A'(t) =0$. Thus the mean age is constant. If we instead want to compute the median age of cited articles, we must solve the equation $P(t-a) = P(t)/2$ for $a$. In the exponential growth phase, we obtain $a = (\log 2)/k$. This is independent of $t$.

%In the current subexponential phase of growth of $P(t)$, it is clear without detailed computations that the average age of cited articles should increase somewhat under our model, since the new articles are crowding out the older ones less quickly and no aging is assumed.

%It is difficult to obtain a precise analytic expression.

\begin{figure}[htbp]
     \centering
     \begin{subfigure}[b]{0.45\textwidth}
         \centering
         \includegraphics[width=\textwidth]{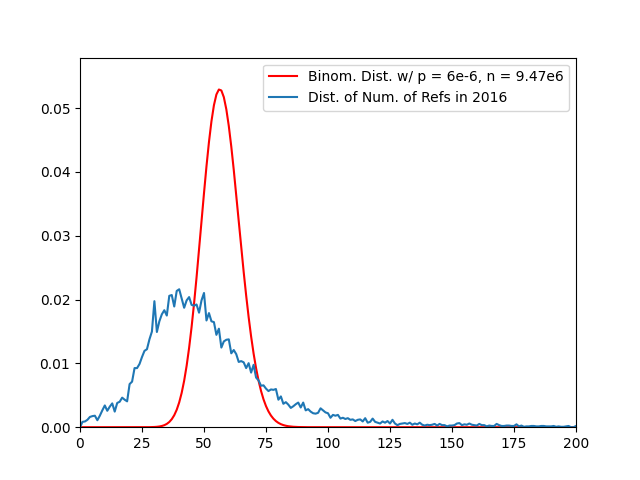}
         \caption{Chemistry}
         \label{fig:Binom_fit_mean_2016_chem}
     \end{subfigure}
     \begin{subfigure}[b]{0.45\textwidth}
         \centering
         \includegraphics[width=\textwidth]{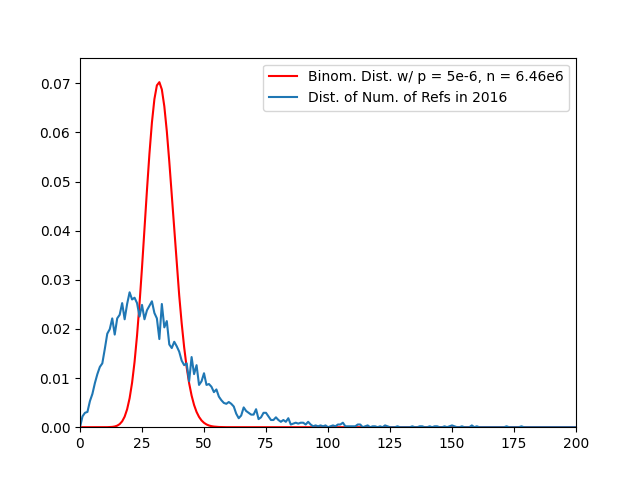}
         \caption{Mathematics}
         \label{fig:Binom_fit_mean_2016_math}
     \end{subfigure}
     \begin{subfigure}[b]{0.45\textwidth}
         \centering
         \includegraphics[width=\textwidth]{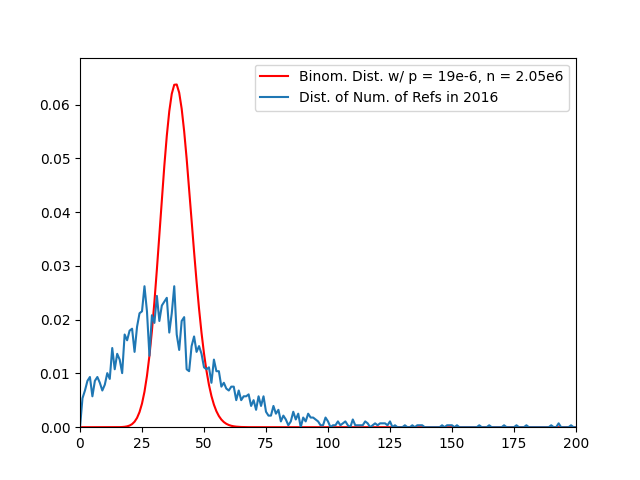}
         \caption{Economics}
         \label{fig:Binom_fit_mean_2016_econ}
     \end{subfigure}
     \begin{subfigure}[b]{0.45\textwidth}
         \centering
         \includegraphics[width=\textwidth]{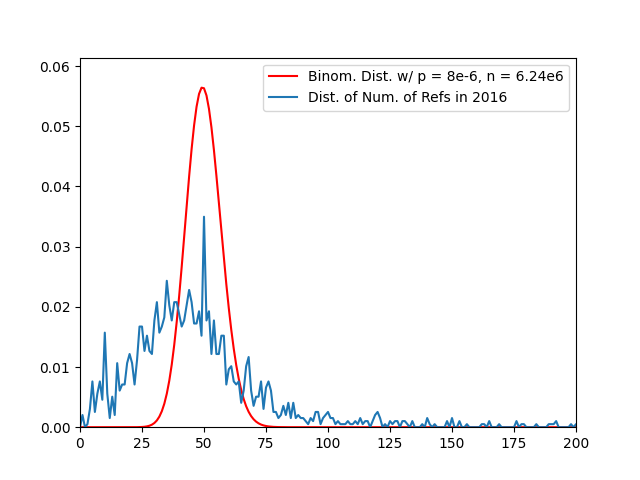}
         \caption{Oncology}
         \label{fig:Binom_fit_mean_2016_onc}
     \end{subfigure}
        \caption{Binomial fitting of mean reference list length of documents published in 2016.}
        \label{fig:Binom_fit_mean_2016}
\end{figure}

It is not to be expected that our mean-field approach should lead to accurate predictions about the \emph{distribution} of the length of the reference list for a randomly chosen article published at time $t$. The model predicts that this should follow a binomial distribution $Bin(P(t), c)$. 
\if Given that $c$ is small, $P(t)$ large and their product in the range 0-100, it makes sense to use a Poisson approximation. 
\fi
However we do consider this more stringent test of the model in Figure~\ref{fig:Binom_fit_mean_2016}. The results are not especially good, as expected, and more detailed modeling of the distribution of reference list length should be undertaken.
%We note in passing that the number of citations of an article published at time $t_0$ by time $t> t_0$ should also be binomially distributed according to $Bin(P(t)-P(t_0), c)$.

\section{Discussion and conclusion}
\label{s:conclusion}

Our results show that the uniform citation model we have used is good enough to explain the time evolution of the length of reference lists and to relate this to the growth in the number of articles published, at least in the modern era of electronic searching. We conclude that the model is promising for statistical studies of this type, and may be of use in other areas. Our more general model may yield more detailed predictions, but we face the serious problem of how to estimate the obsolescence function $q$. It would be very interesting to study a fast-growing modern field which essentially started during the last 20 years, to see how well our uniform model holds up.

Differences in citation patterns of research fields were already discussed by Price \cite{Pric1965}. Fields such as mathematics seem on the face of it to be more suited to the uniform model, where citability of results, once established, presumably does not decline as quickly as in more fast-moving fields.
We hope to inspire other researchers to investigate citation behavior in enough detail to allow improvements on the model presented here, and to distinguish more between growth and obsolescence. 

The empirical and modeling results above lead to policy questions. The increasing growth in the number of references per article (also called ``citation inflation") is potentially a cause for concern, if it continues long enough. This has been noted by previous authors \cite{ULRS+2014, YiBe1991}. Given a constant cost per unit time for reading, scholars will face increasing difficulties in checking references. For example, our linear models predict an increase in the mean of at least 5 references per decade, and this would likely increase. This points to the need for more efficient machine reading of articles, which in turn requires open access to primary research literature. Another, more radical, option would be for journals to restrict the number of references. This is analogous to applicants for grants or promotions focusing on the most important publications. Of course this would not reduce the work of the author, but it would help the reader, and make it easier to compare citation productivity of researchers over time. Alternatively, reasons for citation could be clarified and references to literature standardized. For example, all courtesy citations (where the author mentions work by others in order to convince reviewers that the author is knowledgeable, but that work is not directly built on by the present paper) could be listed in a standard place where readers could classify them as such, and ignore them if time is short. 

\section*{Funding and/or Conflicts of interests/Competing interests}

The authors declare no conflict of interest. No funding supported this work.

\printbibliography

@article{MaBo2016,
  title={Change of perspective: bibliometrics from the point of view of cited references - a literature overview on approaches to the evaluation of cited references in bibliometrics},
  author={Marx, Werner and Bornmann, Lutz},
  journal={Scientometrics},
  volume={109},
  number={2},
  pages={1397--1415},
  year={2016},
  publisher={Springer}
}

@inproceedings{LaAG2007,
	title = {Long-term patterns in the aging of the scientific literature, 1900–2004},
	booktitle = {Proceedings of the 11th conference of the International Society for Scientometrics and Informetrics ({ISSI}), {Madrid}, {Spain}},
	author = {Larivière, Vincent and Archambault, Éric and Gingras, Yves},
	year = {2007},
	pages = {449--456},
}

@article{ULRS+2014,
	title = {Growth in the number of references in engineering journal papers during the 1972–2013 period},
	volume = {98},
	number = {3},
	journal = {Scientometrics},
	author = {Ucar, Iñaki and López-Fernandino, Felipe and Rodriguez-Ulibarri, Pablo and Sesma-Sanchez, Laura and Urrea-Micó, Veronica and Sevilla, Joaquín},
	year = {2014},
	note = {Publisher: Springer},
	pages = {1855--1864},
}

@article{YiBe1991,
	title = {Number of references in biochemistry and other fields; {A} case study of the {Journal} of {Biological} {Chemistry} throughout 1910–1985},
	volume = {21},
	number = {1},
	journal = {Scientometrics},
	author = {Yitzhaki, Moshe and Ben-Tamar, David},
	year = {1991},
	note = {Publisher: Akadémiai Kiadó, co-published with Springer Science+ Business Media BV …},
	pages = {3--22},
}

@article{Pric1965,
  title={Networks of scientific papers: The pattern of bibliographic references indicates the nature of the scientific research front.},
  author={Price, Derek J. De Solla},
  journal={Science},
  volume={149},
  number={3683},
  pages={510--515},
  year={1965},
  publisher={American Association for the Advancement of Science}
}

@article{LiSa1974,
  title={Progress in documentation:‘Obsolescence’and changes in the use of literature with time},
  author={Line, Maurice B. and Sandison, Alexander},
  journal={Journal of Documentation},
  volume={30},
  number={3},
  pages={283--350},
  year={1974}
}

@article{Gupt1998,
	title = {Growth and obsolescence of literature in theoretical population genetics},
	volume = {42},
	number = {3},
	journal = {Scientometrics},
	author = {Gupta, Brij},
	year = {1998},
	pages = {335-347},
}

\appendix
\section{Datasets}
\label{app:methods}
The following description of the dataset was supplied by Academic Analytics.

We culled the Digital Object Identifiers (DOIs) of journal articles (co)authored by scholars in the Academic Analytics commercial database (AAD; \url{http://www.academicanalytics.com/}) between  between 2007 and 2019. The AAD is composed of an annually updated roster of faculty members employed by 390 American Ph.D. granting universities. Faculty members in the AAD are linked to each CrossRef-DOI journal articles they (co)authored using manual and semi-automated disambiguation. The AAD also contains the academic department affiliation(s) of each faculty member; these departments, in turn, are manually classified into disciplines based on NCES CIP codes (\url{https://nces.ed.gov/ipeds/cipcode/}). We extracted from the AAD the DOI of each journal article (co)authored by scholars whose departments are classified in the following disciplines: Chemistry, Economics, English Language and Literature, Geography, History, Mathematics, and Oncology. Each DOI was entered as a search term in the CrossRef API (\url{https://www.crossref.org/}) to retrieve the length of the reference list and the first and last page numbers, from which the total page count of each article was calculated.

\section{Robustness checks}

\subsection{Median vs mean}
\label{ss:medvmean}

\begin{figure}[htbp]
  \includegraphics[width=0.65\linewidth]{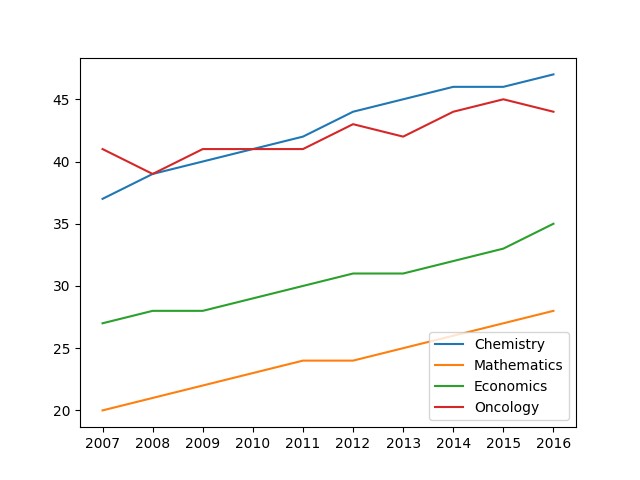}
  \caption{Median reference list length, by year}
  \label{fig:L_median}
\end{figure}

\begin{figure}[htbp]
     \centering
     \begin{subfigure}[b]{0.45\textwidth}
         \centering
         \includegraphics[width=\textwidth]{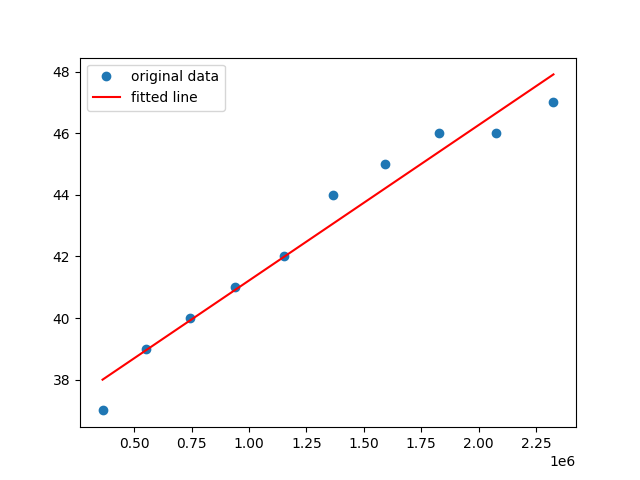}
         \caption{Chemistry -- Slope: $5\times 10^{-6}$}
         \label{fig:LvP_med_chem}
     \end{subfigure}
     \begin{subfigure}[b]{0.45\textwidth}
         \centering
         \includegraphics[width=\textwidth]{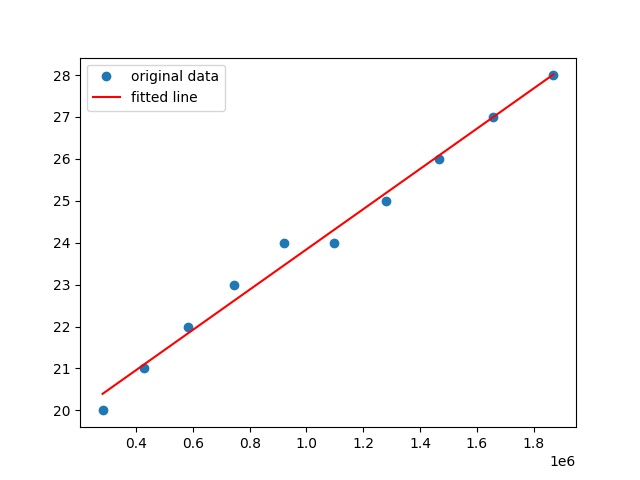}
         \caption{Mathematics -- Slope: $5\times 10^{-6}$}
         \label{fig:LvP_med_math}
     \end{subfigure}
     \begin{subfigure}[b]{0.45\textwidth}
         \centering
         \includegraphics[width=\textwidth]{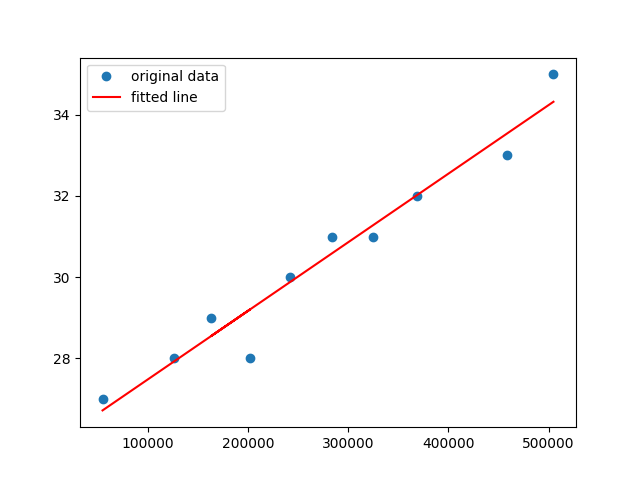}
         \caption{Economics -- Slope: $17\times 10^{-6}$}
         \label{fig:LvP_med_econ}
     \end{subfigure}
     \begin{subfigure}[b]{0.45\textwidth}
         \centering
         \includegraphics[width=\textwidth]{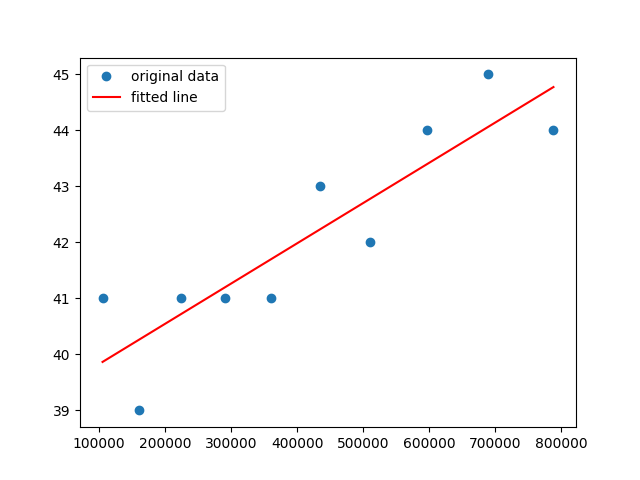}
         \caption{Oncology -- Slope: $7\times{10^{-6}}$}
         \label{fig:LvP_med_onc}
     \end{subfigure}
        \caption{Linear relationship between $P(t)$ and $L(t)$ (Median).}
        \label{fig:LvP_med}
\end{figure}

We repeated the analyses using the mean from the main text, this time using the median. The results were similar, as seen in Figures~\ref{fig:L_median} and ~\ref{fig:LvP_med} and Table~\ref{t:LvP_fit_med}).

\begin{table}[htbp]
    \centering
    \begin{tabular}{c|c}
       Field  & $R^2$ \\
       \hline
       Chemistry  & 0.962511\\
       Mathematics & 0.988723\\
       Economics & 0.980056\\
       Oncology & 0.805837\\
    \end{tabular}
    \caption{Fit of linear model for $L(t)$ versus $P(t)$ - Median}
    \label{t:LvP_fit_med}
\end{table}

\if01
\begin{figure}[htbp]
\includegraphics[width=0.45\linewidth]{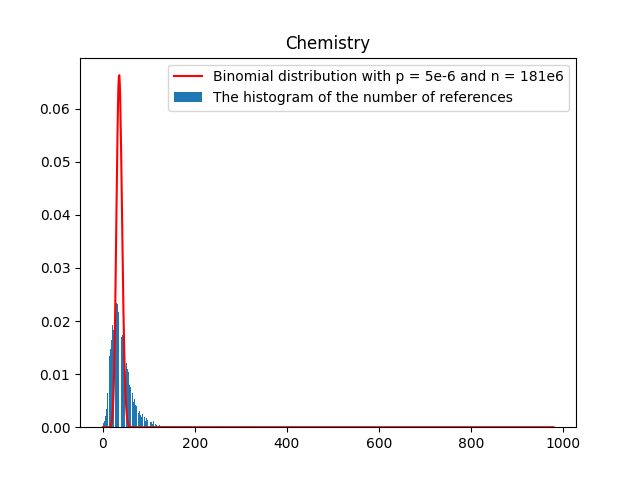}
\includegraphics[width=0.45\linewidth]{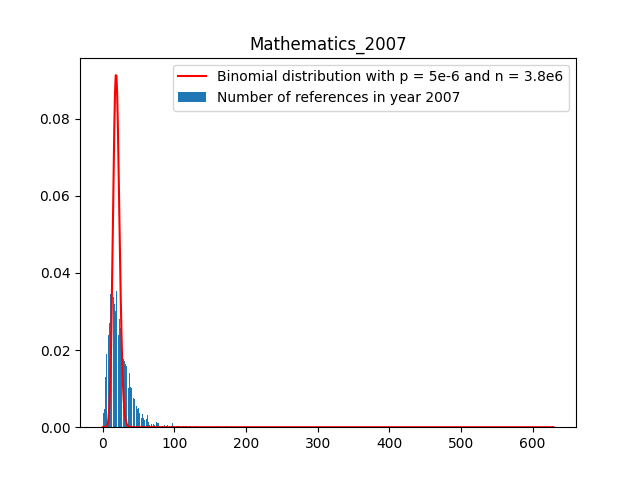}
\includegraphics[width=0.45\linewidth]{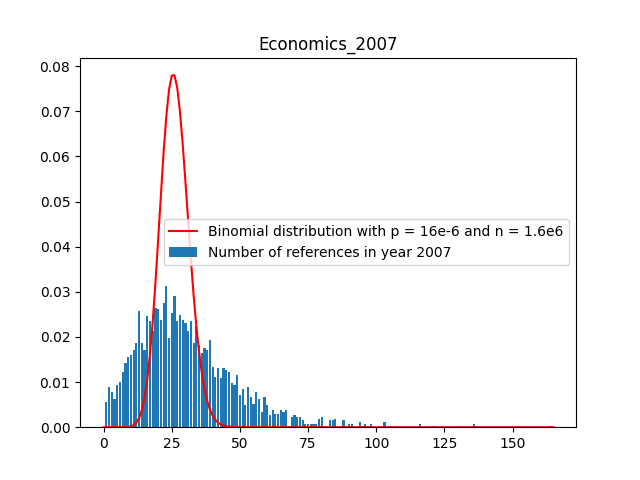}
\includegraphics[width=0.45\linewidth]{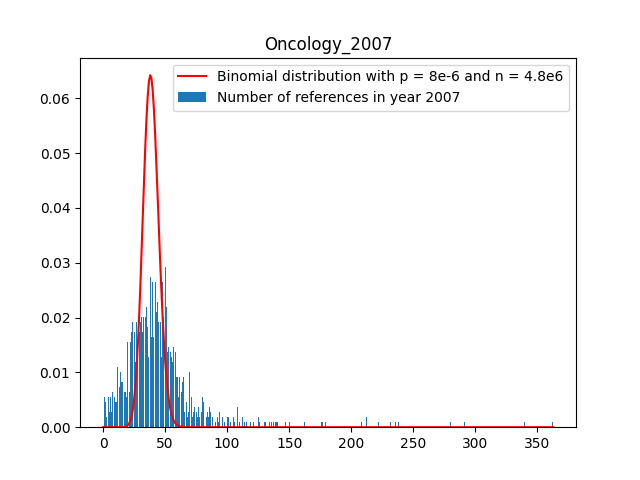}
\caption{Binomial fitting of reference list length - Median}
\label{fig:Binom_fit_med}
\end{figure}
\fi

\end{document}